\documentclass[10pt, twocolumn]{article}

\usepackage{graphicx}
\usepackage{enumerate}
\usepackage[bookmarks=false]{hyperref}

\usepackage{authblk}

\newtheorem{theorem}{Theorem}

\newtheorem{definition}{Definition}
\newtheorem{conjecture}{Conjecture}

\usepackage{bbm}
\usepackage[intoc]{nomencl}
\usepackage{tikz}
\usetikzlibrary{calc,backgrounds}
\usepackage{amsmath,amssymb,amsfonts}
\usepackage[bb=boondox]{mathalfa}

\makeatletter
\newcommand{\LeftEqNo}{\let\veqno\@@leqno}
\makeatother

\title{\LARGE\bf Evolution of Social Power for Opinion Dynamics Networks}
\author[1]{Susana Iglesias Rey
\thanks{Email: \href{mailto:susana.iglesias_rey@nokia.com}{susana.iglesias\_rey@nokia.com}}}
\affil[1]{Nokia Bell Labs\\ Nokia Paris-Saclay\\ Route de Villejust\\ 91620 Nozay\\ France}
\author[2]{Patricio Reyes
\thanks{Email: \href{mailto:patricio.reyes@usc.es}{patricio.reyes@usc.es}}}
\affil[2]{
Technological Institute for Industrial Mathematics   (ITMATI)\\ University of Santiago de Compostela\\ Santiago de Compostela\\ Spain}
\author[1]{Alonso Silva
\thanks{Email: \href{mailto:alonso.silva@nokia-bell-labs.com}{alonso.silva@nokia-bell-labs.com} To whom correspondence should be addressed.}}
\date{}

\begin{document}
\maketitle

\begin{abstract}
This article studies the evolution of opinions and interpersonal influence structures in a group of agents as they discuss a sequence of issues, each of which follows an opinion dynamics model. In this work, we propose a general opinion dynamics model and an evolution of interpersonal influence structures based on the model of \textit{reflected appraisals} proposed by Friedkin. Our contributions can be summarized as follows:
(i) we introduce a model of opinion dynamics and evolution of interpersonal influence structures between issues viewed as a best response cost minimization to the neighbor's actions,
(ii) we show that DeGroot's and Friedkin-Johnsen's models of opinion dynamics and their evolution of interpersonal influence structures are particular cases of our proposed  model, and
(iii) we prove the existence of an equilibrium.
This work is a step towards providing a solid formulation of the evolution of opinions and interpersonal influence structures over a sequence of issues.
\end{abstract}

\section{Introduction}

During the last few years, there has been an increasing interest in understanding how individuals form their opinion. Consequently, a lot of research effort has been devoted to model the underlying process of opinion formation of agents that interact through a social network. In this respect, DeGroot's~\cite{DeGroot74,Berger81} or Friedkin-Johnsen's \cite{FJ99} models are classic references on opinion formation processes.
Other works deal with more general frameworks as nonlinear synchronous models like Krause model \cite{Krause02} or \cite{Lorenz05,Lorenz07}; pairwise asynchronous models \cite{Deffuant00, Frasca13, Frasca15} where opinions take discrete or continuous values, as well as multidimensional \cite{Canuto12,Parsegov17,Parise15, Blondel10} and stochastic \cite{Baccelli14,Baccelli17} models or influences \cite{Varma17}. Usually, these models consider that the payoff of each agent is a function of its own opinion and of an aggregation function, whose value depends on the opinions of an agent-dependent subset of the population, which is defined by the network, usually its neighborhood. Other works include applications of biological systems to social networks, as in \cite{Pluchino05}, or the inclusion of random interactions between agents \cite{Arpan16}. Currently, there is an increasing literature about social networks manipulation, as in \cite{Forster13,Silva17}.
Concurrently with the study of opinion formation, there has been an increasing interest in the literature related to network formation and transformation mechanisms \cite{Skyrms00}, in particular in social networks of agents and their interpersonal influence structures \cite{Friedkin11,Jia15,Jia16}. Within this context, the aim of this article is to study the evolution of opinions and interpersonal influence structures over a sequence of issues in a group of individuals, which in particular generalizes DeGroot's and Friedkin-Johnsen's models. In order to explain this in detail, let us think about a particular issue domain (as politics, statistics, environment, etc.) and a group of people deliberating about a sequence of related issues in that domain (as draft laws, reports, quality evaluation of some new ideas or products, etc.) keeping their relative interpersonal weights fixed and forming their opinions with a particular model of opinion dynamics. This repetitive deliberation across related issues between the same agents  allows us to study the evolution of the group's influence network~\cite{Friedkin11}.\\

As \cite{Jia15} notices, it may be possible to extend any model of opinion dynamics fixed over a single issue, to deal with the evolution of interpersonal influence structures over a sequence of related issues. As we may expect, the model of opinion dynamics proposed will be affected and will affect the model of evolution of the influence network \cite{Jia13}. This idea of interconnection between opinion dynamics and evolution of influence networks is based on the notion of reflected appraisals \cite{Friedkin11, Gecas83}. It is understood as reflected appraisals the psychological phenomenon that individuals' self-appraisals in some dimension (e.g., self-confidence, self-esteem, self-worth) are influenced by the appraisals of other individuals on them. Thus, the social power of an agent can define its self-appraisal, understanding its social power as the control it exerted over the final outcome (the opinions of all the agents in the group over an issue). Regarding this, \cite{Friedkin11}~formalized the evolution of interpersonal influences over a sequence of issues. Recently, quite a few models have been proposed for the evolution of social power. Based on DeGroot's model \cite{DeGroot74} combined with Friedkin's \cite{Friedkin11}, in \cite{Jia13,Jia15} they constructed the \textit{DeGroot-Friedkin model} of opinion dynamics and influence networks. Moreover, \cite{Xu15} proposed the \textit{modified DeGroot-Friedkin model} of opinion dynamics  in order to allow the update of self-appraisal levels in finite time. Last works include the extension of Friedkin-Johnsen's~\cite{FJ99} model to the evolution of social power \cite{Mirtabatabaei14} and the study of the coevolution of appraisal and influence networks \cite{Jia16}.\\

Motivated by these works, we propose a general set of opinion dynamics models, based on \textit{network aggregative games},
which could be extended to deal with the evolution of interpersonal influence structures. In particular, we restrict our attention to opinion formation models in which each agent minimizes a quadratic cost function that depends on its own opinion, a convex combination of the opinions of its neighbors and a convex combination of the initial opinions.

We assume that opinions take values between $0$ and $1$. For example, if agents are discussing about politics, $1$ could be an extremely positive opinion about a particular party and $0$ an extremely negative one. Agents do not want to withdraw from their own opinion and at the same time they do not want to be far away from the opinions of their neighbors. In the interaction model, we incorporate stubbornness of agents with respect to the initial opinions, in the sense that agents do not want to withdraw from their initial opinions. Therefore, at each time step, agents update their opinion minimizing a cost that quantifies the difference between the new opinion and its own current opinion, a convex combination of the opinions of its neighbors and a convex combination of the initial opinions of the agents (it could include, e.g., only its own initial opinion, or its own initial opinion and the initial opinion of its neighbors, but not necessarily be limited to those).

Moreover, each agent updates synchronously its opinion in response to the opinions of its neighbors. We analyze the reflected appraisal mechanism on the evolution of social power for the special case in which the opinion formation process is described by a model included in the general setting proposed, achieving consensus on each of the issues over a sequence. DeGroot's and Friedkin-Johnsen's models of opinion dynamics and their extension to the evolution of the influence network, are particular cases of our setting.\\

Therefore, the main contribution of this paper is the extension of the reflected appraisals dynamics of influence structures for a set of opinion dynamic models, viewed as a myopic cost minimization response to the neighbor's actions (i.e., best response). We construct an equivalent function to the reflected appraisal mechanism on the evolution of social power, proving its continuity and the existence of an equilibrium.\\

This paper is organized as follows. In the next section, we introduce some preliminary notions and the used notation. In Section 3, we propose a general model of opinion dynamics, extension of DeGroot's and Friedkin-Johnsen's models, as the minimization of a quadratic cost. In Section~4, we describe the evolution of  influence structures between issues, defined by a continuous dynamic model. We analyze the existence of an equilibrium and conjecture the existence of convergence. Finally, a discussion of our results and further directions are presented in a closing section.\\

\section{Preliminaries and notation} 

In the following, given a set $S$ the interior of $S$ is denoted by $S^{\mathrm{o}}$. For a column vector $x\in \mathbb{R}^n$, $x^T$ and $\mathrm{diag}(x)$ are used to denote the transposed vector and the diagonal matrix \mbox{$n\times n$} whose diagonal elements are $x_1,x_2,\ldots,x_n$, respectively. The shorthands $\mathbbm{1}_n=(1,\ldots,1)^T$ and $\mathbb{0}_n=(0,\ldots,0)^T$ are adopted. For $i \in \{1,\ldots,n\}$, $e_i\in\mathbb{R}^n$ is the vector with 1 in the \textit{i}th entry and 0 for all other entries. The n-simplex $\Delta_n$ is the set \mbox{$\{x \in \mathbb{R}^n \mid x \geq 0,~\mathbbm{1}_n^T x = 1\}$}. To denote the identity matrix in $\mathbb{R}^{n\times n}$ we use $\mathbb{I}_n$.

A directed graph is a pair $\mathcal{G}=(\mathcal{I},\mathcal{E})$, where $\mathcal{I}$ stands for the finite set of nodes and $\mathcal{E}\subseteq \mathcal{I}\times\mathcal{I}$ is the set of edges. An edge $(i,j)$ will be also denoted as $i \rightarrow j$. A walk from node $i$ to $j$ is a sequence $i\rightarrow i_1 \rightarrow \ldots \rightarrow i_r\rightarrow j$. Node $j$ is \textit{reachable} from node $i$ if there exists a walk from $i$ to $j$. The graph is \textit{strongly connected} if each node is reachable from any other node. A node $i$ is said \textit{globally reachable} if there exist at least one walk from every node $j$ to $i$.

A matrix $M\in \mathbb{R}^{n\times n}$ is said \textit{row stochastic} (analogously, \textit{column stochastic}) if $\sum_{j=1}^{n} m_{ij}=1$ ($\sum_{i=1}^{n} m_{ij}=1$). It is \textit{doubly stochastic} if it is both row and column stochastic. Moreover, $\rho(M)$ denotes the spectral radius of M. Furthermore, $M$ is a \textit{Metzler matrix} if all its non diagonal entries are non-negative. It is \textit{orthogonal} if $M$ is a square matrix with real entries whose columns and rows are orthogonal unit vectors, i.e., $MM^T=M^TM=\mathbb{I}_n$. Finally, given a complex number $z\in\mathbb{C}$, $z=x+iy$, $x, y \in \mathbb{R}$, the \textit{real part} of $z$, denoted by $Re(z)$, is the real number $x$.\\

\section{Model of Opinion dynamics over a network}

Following \cite{Jia13} and \cite{Jia15}, the model of evolution of influence structures is based on two basic ideas. Firstly, we adopt an opinion dynamics model  over a single issue.
Secondly, we adopt the Friedkin's model \cite{Friedkin11} for the dynamics of self-weights and social power over a sequence of issues. Specifically, the first part of the model is defined as follows.\\

\begin{figure}[htb]
\centering
\begin{tikzpicture}[>=latex]
  %
  %
  \tikzstyle{state} = [draw, very thick, fill=gray, circle, minimum height=1.5em, minimum width=1em, node distance=4em]
  \tikzstyle{state2} = [node distance=2em]
  \tikzstyle{stateEdgePortion} = [black,thick];
  \tikzstyle{stateEdge} = [stateEdgePortion,->];
  \tikzstyle{endEdge} = [stateEdgePortion,<-];
  \tikzstyle{edgeLabel} = [pos=0.05,above, text centered];

  %
  %
  \node[state, name=1] {};
  \node[state, name=2, below of=1, right of=1, xshift=2em] {};
  \node[state, name=3, below of=1,  left of=1, xshift=-2em] {i};
  \node[state, name=4, below of=1] {j};
  \node[state, name=5, below of=2, left of=2, xshift=1em ] {};
  \node[state2, name=6, below of=4, xshift=-2em] {(...)};
  \node[state2, name=7, left of=1, xshift=-2em] {(...)};

  %
  %
  \draw (1) 
      edge[endEdge] node[edgeLabel, xshift=-2em]{} 
      (2);
  \draw (1) 
      edge[stateEdge] node[edgeLabel, xshift=-2em]{} 
      (2);
      
    \draw (2) 
        edge[stateEdge] node[edgeLabel, xshift=-2em]{} 
        (5);

  \draw (1) 
      edge[stateEdge] node[edgeLabel, xshift=1em, yshift=2em]{} 
      (3);
      
  \draw (3) 
      edge[stateEdge] node[edgeLabel, xshift=2.5em]{$p_{ij}$} 
      (4);
      
  \draw (4) 
      edge[stateEdge] node[edgeLabel, xshift=-2em]{} 
      (5);
  \draw (4) 
      edge[endEdge] node[edgeLabel, xshift=-2em]{} 
      (5);

  \draw (5) 
      edge[stateEdge] node[edgeLabel, xshift=-2em]{} 
      (6);
      
    \draw (1) 
        edge[stateEdge] node[edgeLabel, xshift=-2em]{} 
        (7);
\end{tikzpicture}
\caption{Scheme of a directed network.}\label{figure1}
\end{figure}
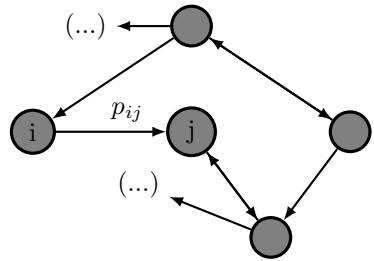

Consider a directed social network $\mathcal{G}=(\mathcal{I},\mathcal{E})$ with $n\geq 2$ agents indexed from $1$ to $n$. Let $\mathcal{I}=\{1,\ldots,n\}$ be the set of agents. Each agent $i\in\mathcal{I}$ has an initial opinion denoted by $y_i\in [0,1]$. Opinions take values between $0$ and $1$. For example, if agents are discussing about the environmental policy of their country, $1$ could be extremely satisfied and $0$~extremely unsatisfied. In the following, the directed network will be specified by the weighted adjacency matrix $P\in\mathbb{R}^{n\times n}$, a row-stochastic  matrix whose diagonal elements are $p_{ii}=0$ and $p_{ij}\in [0,1]$ $i\neq j$. Elements $p_{ij}$ denote the relative weight that agent $i$ gives to the opinion of agent $j$ (see Figure~\ref{figure1}).

For simplicity, we suppose that the graph associated with $P$ is strongly connected. Each agent $i\in\mathcal{I}$ does not want to withdraw from his own opinion and from the opinion of its neighbors. We also incorporate stubbornness of agents with respect to the initial opinions, in the sense that agents do not want to withdraw from their initial opinions. Therefore, each agent has a quadratic cost which wants to minimize that quantify the difference between its own opinion, \linebreak $y_i\in [0,1]$, a convex combination of the initial opinions and a convex combination of the opinions of its neighbors. The convex combination of the opinion of its neighbors, $\sigma_i(t)=\sum_{j\neq i}p_{ij}y_j(t)$, is named the \textit{neighbor's aggregate state} \cite{Parise15} at time $t$. At every time step $t\in\mathbb{N}\cup \{0\}$, each agent $i\in\mathcal{I}$  updates synchronously its opinion minimizing the cost:
\begin{align}
J^i(y,&y_i(t),\sigma_i(t)) =  \label{costo} \\  
&  \alpha y^2 -  2\left(a_iy_i(t)+b_i\sigma_i(t)+\sum_{i=1}^{n}c_{ij}y_j(0)\right)\beta  y + \gamma, \nonumber  
\end{align}
where $a_i\geq 0$, $b_i\geq 0$, $y(0)$ is the vector of opinions at time $t=0$ and  $c_{ij} \in [0, 1]$ are the weights over the vector of initial opinions, for all $i, j \in \mathcal{I}$.

Thus, it is easy to check that the best response strategy for agent $i\in\mathcal{I}$ is given by
\begin{equation}
y_i(t+1)=a_i y_i(t)+b_i\sum_{j\neq i}p_{ij}y_j(t)+\sum_{j=1}^nc_{ij}y_j(0), \label{modelo}
\end{equation}
where the variables $a_i$, $b_i$ and $c_{ij}$ verify
\begin{equation}
a_i + b_i+\sum_{j=1}^nc_{ij}= 1,\quad\forall i\in \mathcal{I}, \label{condicion1}
\end{equation}
to guarantee that the opinions are still in the domain \mbox{$y_i(t)\in [0, 1]$}, for all $ i \in \mathcal{I}$ and for all $t\in\mathbb{N}\cup \{0\}$. This condition and the fact that we establish $C$ as a weight matrix, allows us to define it as the product of a doubly stochastic and orthogonal matrix of weights $Z$, and a diagonal matrix $D$, which ensures that the previous condition is satisfied when $a_i\neq 0$ or $b_i\neq 0$. Therefore, we have that $C=DZ$ and \eqref{condicion1} results in
\begin{equation*}
 a_i + b_i+d_i= 1,\quad\forall i\in \mathcal{I},
\end{equation*}
where $\sum_{j=1}^n z_{ij}=1$, which means that $D=\mathbb{I}_n-A-B$ and $C=(\mathbb{I}_n-A-B)Z$.\\ 

Notice that DeGroot's and Friedkin-Johnsen's models of opinion dynamics can be obtained as particular cases of our cost function:

\begin{itemize}
\item DeGroot's model $y(t+1)=W y(t)$, where $W$ is the network adjacency matrix, is obtained by minimizing the cost given by \eqref{costo} with $a_i=w_{ii}$, $b_i=\frac{1}{1-w_{ii}}$ and $c_{ij}=0$ for all $i, j \in \mathcal{I}$.

\item Analogously, Friedkin-Johnsen's model,  which extends  the  DeGroot's model updating agents opinions including  their prejudices (initial opinions) into  every  iteration, i.e., some  of  the  agents  are stubborn in  the  sense that they are always influenced by their prejudices,
\begin{equation*}
y(t+1)=\varTheta W y(t) + (\mathbb{I}_n-\varTheta)y(0),
\end{equation*}
is obtained minimizing the cost given by \eqref{costo} with $a_i=\theta_i w_{ii}$, $b_i=\frac{\theta_i}{1-w_{ii}}$, $c_{ii}=(1-\theta_i)$ and $c_{ij}=0$ for all $i, j \in \mathcal{I}$, $j\neq i$; where $W$ is the network adjacency matrix and $\varTheta$ a diagonal matrix quantifying the extent to which each individual is open to the influence of its neighbors.\\ 
\end{itemize}

Regarding the evolution of opinion in \eqref{modelo}, in matrix notation, the best-response dynamics are given by
\begin{equation}\LeftEqNo
\tag{I} y(t+1) =\left(A+BP\right)y(t)+Cy(0), \label{modelomatrix}
\end{equation}
where  $y(t)=(y_1(t),\ldots,y_n(t))^T$ is the vector of opinions at time $t$, $A=\mathrm{diag}(a_1,\ldots,a_n)$ and  $B=\mathrm{diag}(b_1,\ldots,b_2)$.\\

The next step to construct the evolution of the influence network is to compute the consensus reached in one issue with the opinion dynamics model \ref{modelomatrix}. In order to compute this consensus, two scenarios have been studied: the case $\sum_{j=1}^nc_{ij}\neq 0$ and the case $\sum_{j=1}^nc_{ij}= 0$.\\

Suppose that  $\sum_{j=1}^nc_{ij}\neq 0$. Then, given condition \eqref{condicion1} and the fact that $P$ is a row stochastic matrix:
\begin{equation*}
\sum_{j\in \mathcal{I} }\left[A+BP\right]_{ij} = a_i + b_i\sum_{j\in \mathcal{I}\backslash\{i\} } p_{ij}=a_i + b_i <1,
\end{equation*}
and consequently $A+BP$ is a row sub-stochastic matrix. Then $\rho\left(A+BP\right)<1$, and we can compute the limit applying the results relative to this kind of matrices,
\begin{align*}
&\lim_{t\rightarrow \infty}y(t) = \lim_{t\rightarrow \infty}\left[\left(A+BP\right)y(t)+Cy(0)\right]=\\
&= \lim_{t\rightarrow \infty}\left[\left(A+BP\right)^ty(0)+ \sum_{k=0}^{t}\left(A+BP\right)^kCy(0)\right]=\\
&= \left(\mathbb{I}_n-(A+BP)\right)^{-1}Cy(0).
\end{align*}
The solution is due to the fact that the first limit satisfies 
\[\lim_{t\rightarrow \infty} \left(A+BP\right)^t=\mathbb{0}_n,\]
and the second one 
\[\lim_{t\rightarrow \infty} \sum_{i=0}^{t}\left(A+BP\right)^t=\left(\mathbb{I}_n-A-BP\right)^{-1},\] see, e.g., Section 5 of \cite{Gebali08}.\\

Therefore, when $\sum_{j=1}^nc_{ij}\neq 0$, consensus takes the form:
\begin{equation*}
y(\infty)=\left(\mathbb{I}_n-A-BP\right)^{-1}Cy(0),
\end{equation*}
which means that agents' opinions converge to a convex combination of their initial opinions weighted by the matrix $\left(\mathbb{I}_n-A-BP\right)^{-1}C$.\\

Let us study the case when $\sum_{j=1}^nc_{ij}= 0$. Then given condition \eqref{condicion1}, we have:
\begin{equation*}
\sum_{j\in \mathcal{I}}\left[A+BP\right]_{ij} = a_i + b_i\sum_{j\in \mathcal{I} \backslash \{i\}} p_{ij}=a_i +b_i \leq 1.
\end{equation*}
If the inequality is strict then $\lim_{t\rightarrow\infty} y(t)=\mathbb{0}_n$ which is a trivial case. Therefore, we are interested in the case:
\begin{equation*}
\sum_{j\in \mathcal{I}}\left[A+BP\right]_{ij} = a_i + b_i\sum_{j\in \mathcal{I} \backslash \{i\}} p_{ij}=a_i +b_i = 1,
\end{equation*}
i.e., $1-a_i=b_i$, $\forall i \in \mathcal{I}$ and therefore $\mathbb{I}_n-A=B$, so the model in \ref{modelomatrix} takes the form:
\begin{equation}\LeftEqNo
 y(t+1)=\left(A+(\mathbb{I}_n-A)P\right)y(t), \tag{II}\label{modelo2}
\end{equation}
where $A+(\mathbb{I}_n-A)P$ is a row stochastic matrix.\\ 

Following the same steps as before, in order to obtain a formulation of the consensus, the limit has been computed
\begin{align*}
&\lim_{t\rightarrow \infty}y(t) = \lim_{t\rightarrow \infty}\left[(A+(\mathbb{I}_n-A)P)y(t)\right]=\\
&\lim_{t\rightarrow \infty}[\left(A+(\mathbb{I}_n-A)P\right)^t]y(0)=v^T y(0) \mathbbm{1}_n.
\end{align*}

The last limit is obtained using Perron-Frobenius theorem~\cite{Gantmacher60} for irreducible matrices, which ensures the existence and uniqueness of $v^T\in \Delta_n$, left eigenvector of $A+(\mathbb{I}_n-A)P$ associated to the eigenvalue $1$. Notice that if the graph associated to $P$ is strongly connected, then $P$ is an irreducible matrix. Therefore, given $P$, row-stochastic irreducible matrix, and $a_i\neq 1$ for all $i \in \mathcal{I}$, we have that $A+(\mathbb{I}_n-A)P$ has the same pattern of zeros and positive entries as $P$, so it is an irreducible matrix too, and we can apply the cited theorem. If $a_i=1$ for some $i \in \mathcal{I}$ then as in Lemma A.1. of~\cite{Jia15}, node $i$ is the only globally reachable node in the graph associated to $A+(\mathbb{I}_n-A)P$, which leads to $v=e_i$. \\

Consequently, when $\sum_{j=1}^nc_{ij}= 0$, consensus takes the form
\begin{equation*}
y(\infty)=v^Ty(0)\mathbbm{1}_n,
\end{equation*}
which means that agents' opinions converge to a convex combination of their initial opinions weighted by the coefficients of $v$.\\

Having a formulation of the model of opinion formation over one single issue and the consensus reached, we can incorporate the evolution of the influence structures over a sequence of related issues.\\


\section{Evolution of social power over a network}

As said before, the model of evolution of influence structures is based on two basic ideas. Firstly, the model of opinion dynamics over a single issue presented in the previous section. Secondly, the Friedkin's model \cite{Friedkin11} for the dynamics of self-weights and power over a sequence of issues. The second part of the model is defined as follows.\\

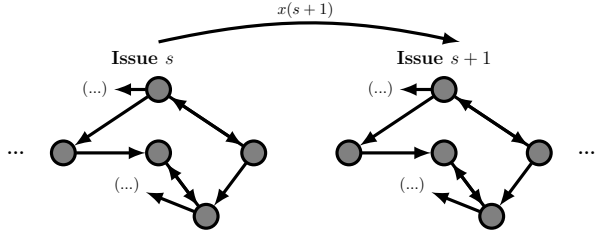
\begin{figure}[htb]
\centering
\begin{tikzpicture}[very thick,scale=0.6, every node/.style={scale=0.6}, >=latex]
  %
  %
  \tikzstyle{state} = [draw, very thick, fill=gray, circle, minimum height=1.5em, minimum width=1em, node distance=4em]
  \tikzstyle{state2} = [node distance=2em]
  \tikzstyle{stateEdgePortion} = [black, very thick];
  \tikzstyle{stateEdge} = [stateEdgePortion,->];
  \tikzstyle{endEdge} = [stateEdgePortion,<-];
  \tikzstyle{edgeLabel} = [pos=0.6,above, text centered];

  %
  %
  \node[state, name=1] {};
  \node[state, name=2, below of=1, right of=1, xshift=2em] {};
  \node[state, name=3, below of=1,  left of=1, xshift=-2em] {};
  \node[state, name=4, below of=1] {};
  \node[state, name=5, below of=2, left of=2, xshift=1em ] {};
  \node[state2, name=6, below of=4, xshift=-2em] {(...)};
  \node[state2, name=7, left of=1, xshift=-2em] {(...)};

  %
  %
  \draw (1) 
      edge[endEdge] node[edgeLabel, xshift=-2em]{} 
      (2);
  \draw (1) 
      edge[stateEdge] node[edgeLabel, xshift=-2em]{} 
      (2);
      
    \draw (2) 
        edge[stateEdge] node[edgeLabel, xshift=-2em]{} 
        (5);

  \draw (1) 
      edge[stateEdge] node[edgeLabel, xshift=1em, yshift=2em]{} 
      (3);
      
  \draw (3) 
      edge[stateEdge] node[edgeLabel, xshift=2.5em]{} 
      (4);
      
  \draw (4) 
      edge[stateEdge] node[edgeLabel, xshift=-2em]{} 
      (5);
  \draw (4) 
      edge[endEdge] node[edgeLabel, xshift=-2em]{} 
      (5);

  \draw (5) 
      edge[stateEdge] node[edgeLabel, xshift=-2em]{} 
      (6);
      
    \draw (1) 
        edge[stateEdge] node[edgeLabel, xshift=-2em]{} 
        (7);

        %
          %
          \node[state, name=12, right of=1, xshift=14em] {};
          \node[state, name=22, below of=12, right of=12, xshift=2em] {};
          \node[state, name=32, below of=12,  left of=12, xshift=-2em] {};
          \node[state, name=42, below of=12] {};
          \node[state, name=52, below of=22, left of=22, xshift=1em ] {};
          \node[state2, name=62, below of=42, xshift=-2em] {(...)};
          \node[state2, name=72, left of=12, xshift=-2em] {(...)};
        
          %
          %
          \draw (12) 
              edge[endEdge] node[edgeLabel, xshift=-2em]{} 
              (22);
          \draw (12) 
              edge[stateEdge] node[edgeLabel, xshift=-2em]{} 
              (22);
              
            \draw (22) 
                edge[stateEdge] node[edgeLabel, xshift=-2em]{} 
                (52);

          \draw (12) 
              edge[stateEdge] node[edgeLabel, xshift=1em, yshift=2em]{} 
              (32);
              
          \draw (32) 
              edge[stateEdge] node[edgeLabel, xshift=2.5em]{} 
              (42);
              
          \draw (42) 
              edge[stateEdge] node[edgeLabel, xshift=-2em]{} 
              (52);
          \draw (42) 
              edge[endEdge] node[edgeLabel, xshift=-2em]{} 
              (52);
        
          \draw (52) 
              edge[stateEdge] node[edgeLabel, xshift=-2em]{} 
              (62);
              
            \draw (12) 
                edge[stateEdge] node[edgeLabel, xshift=-2em]{} 
                (72);

          \node[state2, name=723, right of=22, xshift=1em] {\large \textbf{...}};
          \node[state2, name=724, left of=3, xshift=-1em] {\large \textbf{...}};
          \node[state2, name=issue1, above of=1,xshift=-1em] {\large \textbf{Issue $s$}};
          \node[state2, name=issue2, above of=12] {\large \textbf{Issue $s+1$}};

                    \draw ($(issue1) + (1em,1em)$) 
                        edge[stateEdge, bend left=12.5] node[edgeLabel, xshift=-2em]{$x(s+1)$} 
                        ($(issue2) + (1em,1em)$);
\end{tikzpicture} 
\caption{Scheme of the evolution of self-appraisal.}\label{figure2}
\end{figure}

Consider a fixed directed social network $\mathcal{G}=(\mathcal{I},\mathcal{E})$ with $n\geq 2$ individuals who discuss a sequence of related issues $s\in\mathbb{N}\cup\{0\}$ according to the opinion formation model \ref{modelomatrix}, a relative interaction matrix $P$ and a measure of the self-appraisals $x(s)\in \Delta_n$. Following Friedkin's model \cite{Friedkin11}, in the context of opinion dynamics over a network, we understand as the self-appraisal of agent $i$ the weight that the agent gives to his own opinion. At each issue, agents interact and reach a consensus. Then they update their self-appraisals as a measure of the relative control they had over the consensus \cite{Catwright1959} and discuss the next issue with these updated new weights (Figure~\ref{figure2}).

At fixed issue $s$ and fixed self-weights $x(s)$, the vector of opinions about issue $s$ is a trajectory from time step $t$ to $y(s,t) \in [0, 1]^n$ that evolves according to the extension of model \ref{modelomatrix}. In the following, we assume that the terms $A$, $B$ and $C$ depend on the issue discussed $s\in \mathbb{N}\cup\{0\}$. Specifically, the coefficients of model \ref{modelomatrix} in issue $s+1$ depend on the measure of the previous vector of self-appraisals $x(s)$, i.e., are functions of the vector of self-weights. \\

As in Section 2, we distinguish two models depending on $\sum_{j=1}^nc_{ij}$.\\

If $\sum_{j=1}^nc_{ij}\neq 0$, then the extension of the model in \ref{modelomatrix} in order to include the self-appraisals is:
\begin{equation}
\LeftEqNo y(s,t+1) =(A(x(s))+  B(x(s))P)y(s,t)C (x(s)) y(s,0). \tag{I'}\label{modelo4}
\end{equation}
Analogously, if $\sum_{j=1}^nc_{ij}= 0$ then the extension of model in~\ref{modelo2} is:
\begin{equation}\LeftEqNo
y(s,t+1) =(A(x(s))+(\mathbb{I}_n-A(x(s)))P)y(s,t)  \tag{II'}\label{modelo5}\\
\end{equation}
where matrices $A(x)$ and $B(x)$  are defined as follows. $A(x)$ and $B(x)$ are maps
\begin{align*}
A(\cdot):  \Delta_n &\longrightarrow [0,1]^{n\times n} \\
  (x_1,\ldots,x_n)^T &\longrightarrow \mathrm{diag}(a_1(x_1),\ldots,a_n(x_n)),
\end{align*}
\begin{align*}
B(\cdot):  \Delta_n&\longrightarrow [0,1]^{n\times n} \\
  (x_1,\ldots,x_n)^T &\longrightarrow \mathrm{diag}(b_1(x_1),\ldots,b_n(x_n)),
\end{align*}
where $a_i(x)$ and $b_i(x)$ are analytic functions, for all  $i \in \mathcal{I}$. Moreover, recalling that $C=(\mathbb{I}_n-A-B)Z$, we define $Z(x)$ a doubly stochastic and orthogonal matrix whose elements are continuous functions of $x$ in $[0,1]$.\\

These extensions lead to the consensus outcome of the opinion formation process for model \ref{modelo4} 
\begin{equation*}
y(s,\infty)=\left(\mathbb{I}_n-A(x(s))-B(x(s))P\right)^{-1}C(x(s)) y(s,0),
\end{equation*}
and for model \ref{modelo5}
\begin{equation*}
y(s,\infty)=v(s)^Ty(s,0)\mathbbm{1}_n,
\end{equation*}
where $v(x(s))^T$ is the left eigenvector of \[A(x(s))+(\mathbb{I}_n-A(x(s)))P,\] associated to eigenvalue $1$.\\

As it was said, the reflected-appraisal mechanism is constructed as the relative social power faced by the agent. For social power, we understand the control exercised over the consensus reached \cite{Friedkin11}. This power is measured depending on the consensus reached. Comparing the two presented models, in model \ref{modelo4}, the relative control of each individual is measured via the matrix \[(\mathbb{I}_n-A(x(s))-B(x(s))P)^{-1}C(x(s)),\] thus the evolution of self-appraisals is:
\begin{equation*}
x(s+1)=[\left(\mathbb{I}_n-A(x(s))-B(x(s))P\right)^{-1} C(x(s))]^T  \mathbbm{1}_n/n.  
\end{equation*}

Analogously for model \ref{modelo5}, the relative control of each individual is the corresponding coefficient of $v(x(s))^T$. Thus the vector of self-appraisals is:
\begin{equation*}
x(s+1)=v(x(s)), 
\end{equation*}
where $v(x(s))^T$ is the left eigenvector of the matrix \[A(x(s))+(\mathbb{I}_n-A(x(s)))P,\] associated with eigenvalue $1$. \\

To summarize, we give the following definition:\\

\begin{definition}
Consider a social directed network\break $\mathcal{G}=(\mathcal{I},\mathcal{E})$ with $n\geq 2$ agents discussing a sequence of issues $s\in\mathbb{N}\cup\{0\}$ with relative interaction matrix $P$. The model of evolution of self-appraisals is
\begin{enumerate}
\item for model \ref{modelo4},
\begin{equation}\LeftEqNo\tag{I''}\label{selfapp1}
\begin{gathered}
x(s+1) = \hspace{5.4cm}\\
 [\left(\mathbb{I}_n-A(x(s))-B(x(s))P\right)^{-1}C(x(s))]^T  \mathbbm{1}_n/n.
\end{gathered}
\end{equation}
\item  Analogously, for model \ref{modelo5},
\begin{equation}\LeftEqNo
x(s+1)=v(x(s)), \tag{II''}\label{selfapp2}
\end{equation}
where $v(x(s))^T$ is the left eigenvector of \[A(x(s))+(\mathbb{I}_n-A(x(s)))P,\] associated with eigenvalue $1$.\\
\end{enumerate}
\end{definition}

Next, we present a theorem with equivalent expressions for the evolution of self-appraisals. Notice that for the evolution given in \ref{selfapp1} with $C=DZ$, if $d_i(x_i)=0$ for any $i \in \mathcal{I}$ we have that $x_i(s)=0$ for all $s\in \mathbb{N}\cup\{0\}$. Thus, in the following we are going to assume that $d_i(x_i)\neq 0$ for all $i \in \mathcal{I}$.\\

\begin{theorem}[Evolution of Self-appraisals]
Consider a directed network $\mathcal{G}=(\mathcal{I},\mathcal{E})$ with $n\geq 2$ agents: 
\begin{enumerate}
\item \label{Teorema11} (Generalization of Theorem III.2 in \cite{Jia13})  The process \ref{selfapp1} is equivalent to $x(s+1)=F(x(s))$ where the continuous map $F:\Delta_n\rightarrow\Delta_n$ is  $u(x(s))Z(x(s))$ where $u(x(s))$ is the left eigenvector of: 
\begin{align*}
U&(x(s)) = \mathbbm{1}_n\mathbbm{1}_n^T/n ~ -  \\
&\left[\mathbb{I}_n-A(x(s))-B(x(s))\right]^{-1}B(x(s))\left(\mathbb{I}_n-P\right).
\end{align*}
\item \label{Teorema12} (Generalization of Lemma 2.2. in \cite{Jia15})  Analogously, model \ref{selfapp2} is equivalent to \[x(s+1)=F(x(s)),\] where $F:\Delta_n\rightarrow\Delta_n $ is a continuous map defined by:
\begin{equation*}
F(x)= 
\begin{cases} 
      e_i, \hspace{2cm} \text{if } a_i(x_i)=1 \text{ for all } i\in \mathcal{I}, \\
     \resizebox{0.28\textwidth}{!}{$\displaystyle\left(\frac{p_1}{1-a_1(x_1)},\ldots,\frac{p_n}{1-a_n(x_n)}\right)^T/\sum_{i\in I}\frac{p_i}{1-a_i(x_i)}$}, ~~~ \text{o/w},
\end{cases} 
\end{equation*}
where $p^T$ is the dominant left eigenvector of the relative interaction matrix $P$. \\
\end{enumerate}
 
\end{theorem}

\textit{Proof.}
For statement \ref{Teorema11}, given the decomposition $C=(\mathbb{I}_n-A-B)Z$, the evolution of self-appraisals~\ref{selfapp1} can be rewritten as:
\begin{align*}
x&(s+1)^T\cdot\\
&\left[\left(Id-A(x(s))-B(x(s))P\right)^{-1} C (x(s)) \right]^{-1} = \mathbbm{1}_n^T/n \Leftrightarrow \\
x&(s+1)^T\cdot\\
&\left[C (x(s))^{-1}\left(Id-A(x(s))-B(x(s))P\right)\right] = \mathbbm{1}_n^T/n. 
\end{align*}
Then,
\begin{align*}
x(s+1)^TZ(x(s))^T&\left[\mathbb{I}_n-A(x(s))-B(x(s))\right]^{-1}\cdot\\
&\left(\mathbb{I}_n-A(x(s))-B(x(s))P\right) = \mathbbm{1}_n^T/n,
\end{align*}
taking into account that $Z$ is an orthogonal matrix and the inverse of an orthogonal matrix is its transpose.\footnote{Notice that another possibility is to define $Z$ as a matrix that satisfies that the product of a vector in $\Delta_n$ and $Z^{-1}$ has positive elements. Then the result would be equivalent but with $Z(x(s))^{-1}$ instead of $Z(x(s))^T$.}

Moreover, the product,
\begin{align*}
[\mathbb{I}_n-A(x(s &))-B(x(s))]^{-1}\\
&\cdot (\mathbb{I}_n-A(x(s))-B(x(s))P) = \mathbbm{1}_n^T/n,
\end{align*}
can be rewritten as:
\begin{equation*}
\mathbb{I}_n+[\mathbb{I}_n-A(x(s))-B(x(s))]^{-1} B(x(s))(\mathbb{I}_n-P). 
\end{equation*}
Then,
\begin{align*}
x(s+1&)^TZ(x(s))^T = \mathbbm{1}_n^T/n  + x(s+1)^TZ(x(s))^T\cdot\\
&\left[\mathbb{I}_n-A(x(s))-B(x(s))\right]^{-1}B(x(s))\left(\mathbb{I}_n-P\right).
\end{align*}

$Z(x(s))$ is a doubly stochastic matrix and therefore  $Z(x(s))^T$ is also doubly stochastic. Moreover, $x(s+1)\in \Delta_n$, so we have that 
\[x(s+1)^TZ(x(s))^T\mathbbm{1}_n\mathbbm{1}_n^T=\mathbbm{1}_n^T.\]
Due to this, \[x(s+1)^TZ(x(s))^T=x(s+1)^TZ(x(s))^TU(x(s)),\] therefore $x(s+1)^TZ(x(s))^T$ is a left eigenvector of $U(x(s))$. Moreover, $x(s+1)^TZ(x(s))^T$ has all its entries positive because $x(s+1)\in \Delta_n$ and $Z(x(s))^T$ is a stochastic matrix. It remains to prove that it is the only left eigenvector of $U(x(s))$ with such characteristic. 

Notice that $U(x)$ is a row stochastic matrix, therefore given Perron-Frobenius theorem \cite{Gantmacher60},  it exists a left eigenvector associated to eigenvalue $1$.

Moreover, using Gershgorin's discs theorem \cite{Richard94} and following the proof of Theorem III.2 in \cite{Mirtabatabaei14}, the eigenvectors of $U(x)$ should be at least in one of the Gershorin's discs. In our case these discs are characterized by its centers and radius:\\

\begin{itemize}
\item centers: $u_{ii}(s) = \frac{1}{n} - \frac{b_i(x_i(s))}{1-a_i(x_i(s))-b_i(x_i(s))}$.\\
\item radius: $\sum_{j\neq i} u_{ij}(s)= \frac{n-1}{n} + \frac{b_i(x_i(s))}{1-a_i(x_i(s))-b_i(x_i(s))}$. \\
\end{itemize}

Therefore, all the discs are at the left side of $Re(\lambda)=1$ on the complex plane. This means that the real parts of $U(x(s))$ eigenvectors are less or equal than $1$. Moreover  $U(x(s))$ is a Metzler matrix. Consequently,  
it exists $\tau$ so that the entries of $U(x(s))+\tau \mathbb{I}_n$ are strictly positive. Using Perron-Frobenius theorem \cite{Gantmacher60} for positive matrices, $\rho(U(x(s))+\tau \mathbb{I}_n)$ is a simple real eigenvalue and it exists an associated eigenvector whose entries are strictly positive. Therefore, $U(x(s))$ has a simple real eigenvalue equal to $\rho(U(x(s))+\tau \mathbb{I}_n)-\tau$, that according to the previous discussion is equal to $1$, with a unique eigenvector $u(x(s))\in \Delta_n^{\mathrm{o}}$.\\

For statement \ref{Teorema12}, following \cite{Jia15}, we are interested in the equality:
\begin{equation*}
[A(x(s))+(\mathbb{I}_n-A(x(s)))P]^Tx(s+1)=x(s+1),
\end{equation*}
which can be rewritten as,
\begin{equation*}
[\mathbb{I}_n-A(x(s))-P^T(\mathbb{I}_n-A(x(s)))]x(s+1) = 0_n.
\end{equation*}
Moreover,
\begin{equation*}
(\mathbb{I}_n-P^T)(\mathbb{I}_n-A(x(s)))x(s+1)=0_n.
\end{equation*}
Then we have that,
\begin{equation*}
[\mathbb{I}_n-A(x(s))]x(s+1)=P^T[\mathbb{I}_n-A(x(s))]x(s+1).
\end{equation*}
Therefore $x(s+1)^T[\mathbb{I}_n-A(x(s))]^T$ is the left eigenvector of  $P$ associated with eigenvalue $1$. Then, \[[\mathbb{I}_n-A(x(s))]x(s+1)=\alpha(s)p,\] where $\alpha(s)=1/\sum_{i=1}^{n}\frac{p_i}{1-a_i(x_i(s))}$ is a scale factor in order to guarantee that $x(s+1)\in\Delta_n$. Consequently, we have that:
\begin{equation*}
x(s+1)=\alpha(s)\left(\frac{p_1}{1-a_1(x_1(s))},\ldots,\frac{p_n}{1-a_n(x_n(s))}\right)^T.
\end{equation*}

It is easy to see that $F(x)$ is the composition of functions $G(x)$ and $A(x)$ where:
\[ G(x)=\left(\frac{p_1}{1-x_1},\ldots,\frac{p_n}{1-x_n}\right)^T/\sum_{i=1}^{n}\frac{p_i}{1-x_i}, 
\]
is an analytic function in $\Delta_n\backslash\{e_1,\ldots,e_n\}$ and hence, continuous in $\Delta_n\backslash\{e_1,\ldots,e_n\}$. By definition, $A(x)$ is an analytic function and hence is continuous, which means that the map $F(x)=(G\circ A)(x)$ is a composition of continuous functions in $\Delta_n\backslash\{e_1,\ldots,e_n\}$, i.e., it is continuous in $\Delta_n\backslash\{e_1,\ldots,e_n\}$.\\

The rest of the proof of the continuity of $F$ can be found in Lemma 2.2. of \cite{Jia15}. $\Box$\\

The characterization of the evolution of the self-appraisals provided in the previous theorem allows us to formulate the following result, generalization of the work of \cite{Mirtabatabaei14} and \cite{Jia15}.\\

\begin{theorem}[Equilibrium] Given a directed network \break $\mathcal{G}=(\mathcal{I},\mathcal{E})$ with $n\geq 3$ agents:

\begin{enumerate}
\item \label{Teorema21} Consider the dynamical system in Theorem 1.\ref{Teorema11} and assume that $d_i(x_i)\neq 0$ for all $i \in I$. Then,  it exists $x^*$ equilibrium point of \ref{selfapp1} that belongs to the interior of $\Delta_n$.
\item \label{Teorema22}Assume the digraph associated to P does not have a star topology and consider the dynamical system in Theorem 1.\ref{Teorema12}. Then, it exists $x^*$ equilibrium point of \ref{selfapp2} that belongs to the interior of $\Delta_n$.
Moreover, if $A(x)=x$ then $x^*$ is unique, the ordering of the entries of $x^*$ is equal to the ordering of $p$, and $\{e_1,\ldots,e_n\}$ are the only equilibrium points of $F$ on the boundary of $\Delta_n$~\cite{Jia15}.\\
\end{enumerate}
\end{theorem}

\textit{Proof.}
For statement \ref{Teorema21}, notice that according to Theorem~1.~\ref{Teorema11} the evolution of self-appraisals can be represented by a function $F$ that maps any self-appraisal vector \mbox{$x \in \Delta_n$} to the product of the unique positive dominant left eigenvector of $U(x)$ and $Z(x)$. Following Lemma III.3 of \cite{Mirtabatabaei14} it can be proved that if $M(t)$ is a square matrix whose entries are analytic functions of parameter $t$ then it exists an eigenvector whose entries are real analytic functions of $t$.
The entries of matrix $U(x)$ are analytic functions. $A$ and $B$ were defined as analytic functions and notice that the sums and products of analytic functions and the inverse of an analytic function whose derivative is nowhere zero are analytic. Consequently, the entries of $u(x)$ are real analytic functions of $x$. Moreover, by definition, the entries of $Z(x)$ are continuous functions of~$x$. Thus $F(x)$ is a continuous function of $x$.

According to Brouwer fixed-point theorem, because $F(x)$ is a continuous function that maps a convex compact subset of an Euclidean space into itself, it has at least one fixed point $x^* \in \Delta_n^{\mathrm{o}}$. \\

For statement \ref{Teorema22}, the evolution of self-appraisals can be represented by a particular function $F$. We can show, as in Theorem 4.1. in \cite{Jia15}, that $F$ is a continuous map on a compact set, then the Brouwer fixed-point theorem implies the existence of at least one fixed point on the compact set. Moreover, if we take $A(x)=x$ the model is equal to \cite{Jia15} and hence the statement. $\Box$\\

Previous results (\cite{Mirtabatabaei14, Jia13, Jia15}) illustrate that for some values of $A$, $B$ and $C$ and network topologies, the trajectories converge. Therefore, we formulate the next conjecture that remains to be proven in future work.\\
\begin{conjecture}
For  the dynamical systems of the evolution of social power \ref{selfapp1} and \ref{selfapp2} with $a_i\neq 1$ for all $i\in \mathcal{I}$, the vector of self-appraisals $x(s)\in\Delta_n$ converges to an equilibrium configuration as $s \rightarrow \infty$.\\
\end{conjecture}

In order to check the previous conjecture we simulate a small example of a strongly connected network with three agents that interact through a particular opinion dynamics model of \ref{modelomatrix}. Selecting different maps for $A$, $B$ and $C$ and different initial vectors of self-appraisals $x(0)\in \Delta_n$, we simulate the evolution of the social power.

\begin{figure}[htb!]
\centering
\includegraphics[width=0.38\textwidth]{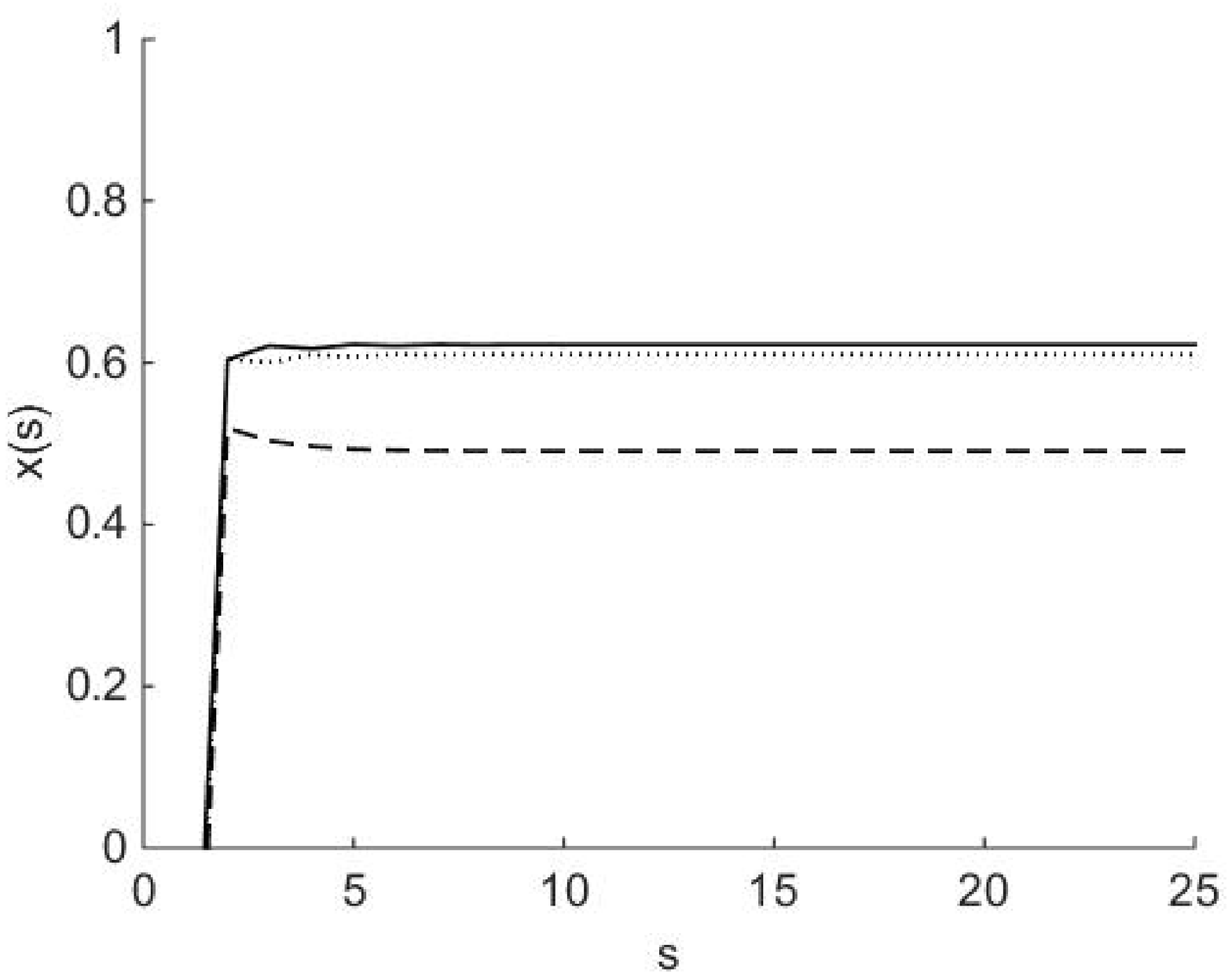}\\
\includegraphics[width=0.38\textwidth]{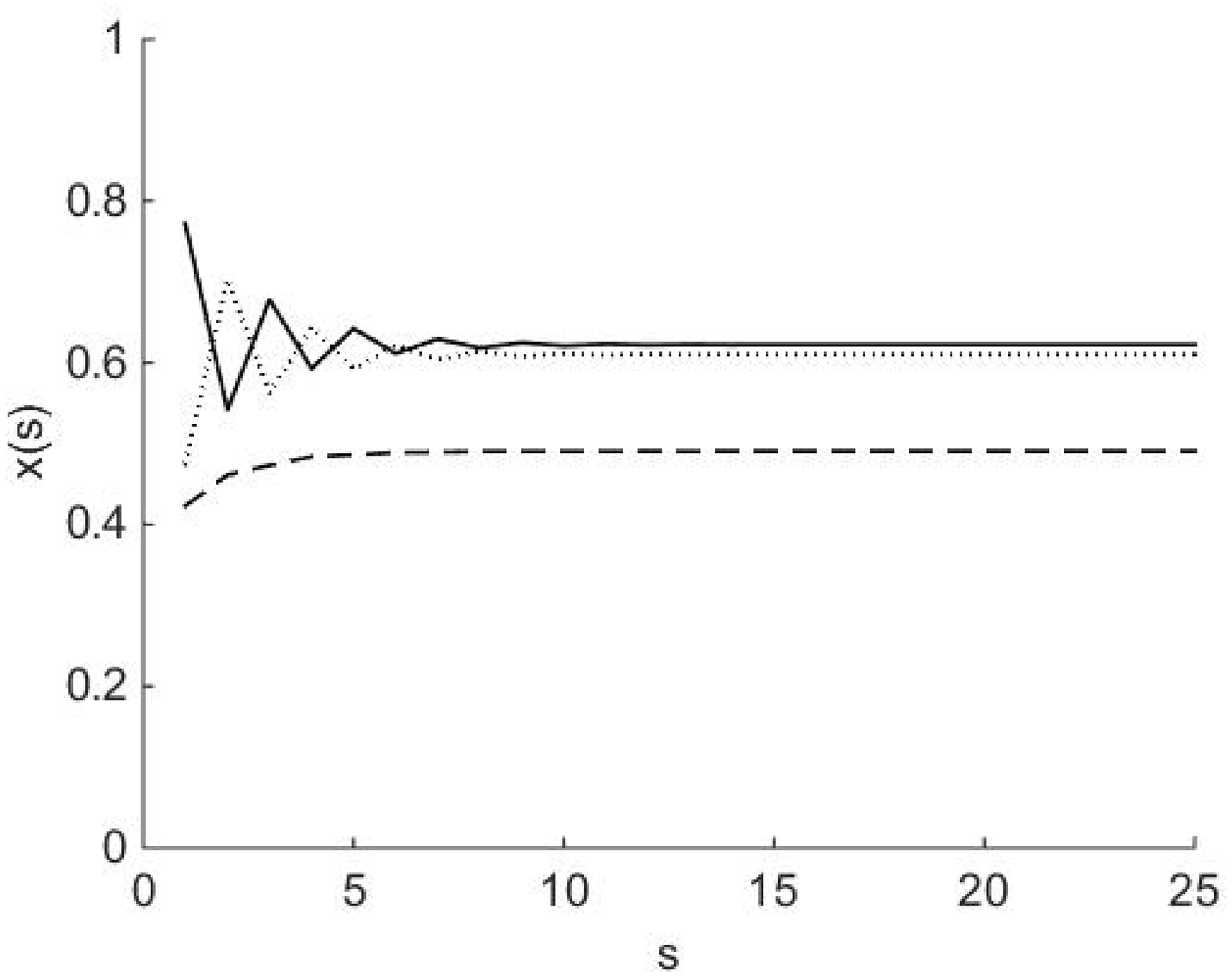}
\caption{Evolution of self-appraisals with $n=3$ agents, different variables and initial opinions.}
\label{Conv1}
\end{figure}

We notice that our simulations (Figure \ref{Conv1}) also show the convergence of the systems of the evolution of self-appraisals. Moreover, the values become stable after a few interactions.\\

\section{Conclusions}

In this paper, we have introduced a general model of opinion dynamics in which each agent minimizes a quadratic cost function that depends on its own opinion, a convex combination of the opinions of its neighbors and a convex combination of the agents' initial opinions (it could include, e.g., only its own initial opinion or its own initial opinion and the initial opinion of its neighbors but not necessarily be limited to those). We have analyzed the evolution of interpersonal influence structures in a group that discuss a sequence of related issues inspired by Friedkin's reflected appraisal mechanism. We have developed a formulation of the dynamical system for the evolution of self-appraisals which is a generalization of the models of \cite{Jia13,Jia15} and proved the existence of a fixed point in the interpersonal influence structures. Many questions still remain to be answered. Future work would be devoted to demonstrate the uniqueness of the equilibrium and the convergence of the evolution of self-appraisals. This will lead to the mathematical characterization of individuals final self-appraisal and their dependence on the system's parameters.\\

\section*{Acknowledgements}
The work of A.~Silva and S.~Iglesias Rey was partially carried out at LINCS (\url{www.lincs.fr}).

\bibliographystyle{hieeetr}
\bibliography{bibliography}

\end{document}